\documentclass[10pt,twocolumn,superscriptaddress,groupedaddress,nofootinbib,aps,prd,preprintnumbers]{revtex4-1}

\usepackage{graphicx}
\usepackage{bbm}
\usepackage{amsmath, amssymb}

\usepackage{hyperref}

\newcommand{\IP}{\mathbb{P}}

\newcommand{\IZ}{\mathbb{Z}}
\newcommand{\cO}{{\cal O}}

\newcommand{\cC}{{\cal C}}

\newcommand{\cK}{{\cal K}}

\newenvironment{packed_enum}{
\begin{enumerate}
 \setlength{\topsep}{0pt} 
  \setlength{\itemsep}{0pt}
  \setlength{\parskip}{0pt}
  \setlength{\parsep}{0pt}
  \setlength{\partopsep}{0pt}  
}{\end{enumerate}}

\begin{document}

\preprint{CERN-TH-2021-051}

\title{Swampland Conjectures and Infinite Flop Chains}
\author{Callum~R.~Brodie}
\email[]{callum.brodie@ipht.fr}
\affiliation{Institut de Physique Th\'{e}orique, Universit\'{e} Paris Saclay, CEA, CNRS Orme des Merisiers, 91191 Gif-sur-Yvette CEDEX, France}
\author{Andrei Constantin}
\email[]{andrei.constantin@physics.ox.ac.uk}
\affiliation{Rudolf Peierls Centre for Theoretical Physics, University of Oxford, Parks Road, Oxford OX1 3PU, UK}
\affiliation{Mansfield College, University of Oxford, Mansfield Road, OX1 3TF, UK}
\author{Andre Lukas}
\email[]{andre.lukas@physics.ox.ac.uk}
\affiliation{Rudolf Peierls Centre for Theoretical Physics, University of Oxford, Parks Road, Oxford OX1 3PU, UK}
\author{Fabian Ruehle}
\email[]{fabian.ruehle@cern.ch}
\affiliation{CERN Theory Department, 1 Esplanade des Particules, CH-1211 Geneva, Switzerland}
\affiliation{Rudolf Peierls Centre for Theoretical Physics, University of Oxford, Parks Road, Oxford OX1 3PU, UK} 

\begin{abstract}\noindent
We investigate swampland conjectures for quantum gravity in the context of M-theory compactified on Calabi-Yau threefolds which admit infinite sequences of flops. Naively, the moduli space of such compactifications contains paths of arbitrary geodesic length traversing an arbitrarily large number of K\"ahler cones, along which the low-energy spectrum remains virtually unchanged. In cases where the infinite chain of Calabi-Yau manifolds involves only a finite number of isomorphism classes, the moduli space has an infinite discrete symmetry which relates the isomorphic manifolds connected by flops. This  is a remnant of the 11D Poincare symmetry and consequently gauged, as it has to be by the no-global symmetry conjecture. The apparent contradiction with the swampland distance conjecture is hence resolved after dividing by this discrete symmetry. If the flop sequence involves infinitely many non-isomorphic manifolds, this resolution is no longer available. However, such a situation cannot occur if the Kawamata-Morrison conjecture for Calabi-Yau threefolds is true. Conversely, the swampland distance conjecture, when applied to infinite flop chains, implies the Kawamata-Morrison conjecture under a plausible assumption on the diameter of the K\"ahler cones.
\end{abstract}

\maketitle

\section{Introduction}
\noindent
When studying string-derived models for particle physics and cosmology, it is an important question which features are universally present and which ones are universally excluded. Such model-independent features may be general enough to persist beyond string theory, thus making predictions for other UV-complete theories of gravity. This line of thought has led, especially in recent years, to a wealth of quantum gravity conjectures (see Ref.~\cite{Palti:2019pca} for a review). One of the oldest and most studied such conjectures states that any global symmetry in quantum gravity must be either gauged or broken \cite{Misner:1957mt, Banks:2010zn}.  Hence a global unbroken symmetry is a feature thought to never occur in quantum gravity. An example of a feature thought to be universal is captured by the swampland distance conjecture~\cite{Ooguri:2006in}, which states that, compared to a theory at a point $p_0$ in moduli space, a theory at point $p_1$ with shortest geodesic distance $\Delta\tau$ from $p_0$ has an infinite tower of light states starting with a mass of the order $e^{-\rho \Delta\tau}$ for some order one constant $\rho>0$. The conjecture has been verified asymptotically in several examples~\cite{Grimm:2018ohb,Blumenhagen:2018nts,Joshi:2019nzi,Erkinger:2019umg,Lee:2019wij,Klaewer:2020lfg, Enriquez-Rojo:2020pqm} (see~table 3 of Ref.~\cite{Andriot:2020lea} for a nice overview), as well as numerically in a recent study based on numerical CY metrics~\cite{Ashmore:2021qdf}. 

A ubiquitous feature of string compactifications (and, possibly, of quantum gravity more generally) are topology-changing transitions, such as flop and conifold transitions. It is an interesting question whether these transitions might imply universal properties of infrared theories. In this letter we will investigate their relationship to existing quantum gravity conjectures.

More concretely, we will study the relation between flop transitions, in the context of M-theory on Calabi-Yau (CY) threefolds, and the swampland distance conjecture (see Ref.~\cite{Heidenreich:2020ptx} for a recent discussion of swampland conjectures in 5D M-theory).

Flop transitions in string theory have been studied some time ago (see Refs.~\cite{Aspinwall:1993nu, Strominger:1995cz, Greene:1995hu, Witten:1993yc}, and Ref.~\cite{Greene:1996cy} for a review) but recently a number of interesting features have been observed~\cite{CICYFlops, Brodie:2020fiq}:
\begin{packed_enum}
 \item[$\bullet$] Many of the simplest CY threefolds, including complete intersections in products of projective spaces (CICYs), admit flop transitions, often in multiple directions in the K\"ahler cone.
 \item[$\bullet$] Flops frequently connect isomorphic manifolds and in this case their moduli spaces are related by a discrete symmetry.
 \item[$\bullet$] Infinite chains of flops arise frequently and even for relatively simple manifolds. 
\end{packed_enum}

It is the last of these features which is the impetus for this present study of the swampland distance conjecture. Infinite chains of flop transitions seem to imply the existence of infinite length geodesics, traversing an arbitrary number of K\"ahler cones but with the low-energy spectrum virtually unchanged. Naively, this appears to contradict the swampland distance conjecture. A priori, one can conceive of two qualitatively different types of infinite flop chains:
\begin{packed_enum}
\item[(1)] The chain only contains a finite number of non-isomorphic CY threefolds.
\item[(2)] The chain contains an infinite number of non-isomorphic CY threefolds.
\end{packed_enum}
In this letter, we would like to make three main points. First, we will show that there indeed exist infinite-length geodesics along which the low-energy spectrum is virtually unchanged for CY threefolds with infinite flop chains. Second, for infinite chains of type (1) we establish the existence of an infinite discrete symmetry on the moduli space which must be gauged, and determine the source of this gauging. Dividing by this symmetry leaves only a finite number of inequivalent K\"ahler cones and this removes any possible conflict with the swampland distance conjecture. This argument does not apply to infinite flop sequences of type (2). As our third main point, we argue that such cases are in fact excluded, provided the Kawamata-Morrison conjecture for CY threefolds holds.

The rest of the paper is organized as follows. In the next section we provide a brief summary of background material, including CY flops and infinite flop chains, the Kawamata-Morrison conjecture and M-theory compactification of CY threefolds. In Section~\ref{sec:Geodesics} we study how geodesics traversing flop transitions relate to the distance conjecture and we conclude in Section~\ref{sec:conclusion}.

\section{Background}
\subsection{Flops}\noindent
Flops can be discussed in generality in the context of algebraic varieties, however some ideas become more transparent in the K\"ahler setting. Thus let $X$ denote a K\"ahler threefold and $J$ its K\"ahler form. The requirement that $J$ is a positive $(1,1)$-form defines the K\"ahler cone $\cK(X)$.

The closure $\bar{\mathcal{K}}(X)$ of the K\"ahler cone is the nef cone. At the boundaries of the K\"ahler cone the manifold becomes singular due to the vanishing of the volume of either a curve $C$, a divisor $D$, or $X$ itself, where
\begin{equation*}
{\rm vol }(C)=\int_C J,\, ~ {\rm vol }(D)=\frac{1}{2}\int_D J^2,\, ~ {\rm vol }(X)=\frac{1}{6}\int_X J^3\, .
\end{equation*}

A flop is a birational morphism. It relates two manifolds by contracting curves on either manifold to arrive at the same (singular) manifold at a common boundary of their K\"ahler cones. A flop can be thought of as a codimension-two surgery, in which a finite collection of isolated rational curves on~$X$ is replaced by another finite collection of isolated rational curves to give rise to a new manifold $X'$. Since divisors are codimension-one objects, there is a natural identification $H^2(X,\mathbb R)\cong H^2(X', \mathbb R)$ and we can speak of the K\"ahler cones $\cK(X)$ and $\cK(X')$ as cones in the same vector space of real divisor classes, as depicted in Fig.~\ref{fig:extendedKCsym} and Fig.~\ref{fig:extendedKCinf} below. The union of the K\"ahler cones of all threefolds obtained from $X$ through a sequence of flops is called the {\itshape extended K\"ahler cone} of $X$ and denoted by $\mathcal{K}_{\rm ext}(X)$.

If $X$ is a CY threefold and $X\rightarrow X'$ a flop, the threefold $X'$ is also CY and has the same Hodge numbers as $X$. However, finer topological invariants, such as the intersection numbers and the second Chern class of the manifold, do change in general. 

Concretely, if $D$ is a divisor on $X$ and $D'$ the corresponding divisor on $X'$, the triple self-intersection form on $H^2(X,\mathbb Z)$ changes as
\begin{equation}\label{eq:top_data_change}
D'^3 = D^3 - \sum_{i=1}^N (D\cdot \cC_i)^3 \; ,
\end{equation}
where $\cC_1,\cC_2,\ldots,\cC_N$ are the isolated exceptional $\IP^1$ curves with normal bundle $\cO(-1)\oplus\cO(-1)$ contracted in the flop. 
Often, the flopping curves belong to a single homology class $\eta\in H_2(X,\mathbb Z)$. There are also cases where some curves belong to the primitive class $\eta$ and the rest to $2\eta$. The numbers of rational curves in $\eta$ and $2\eta$ are the genus-zero Gromov-Witten invariants for these classes. The class $\eta$ is perpendicular, with respect to the intersection form on $X$, to the wall separating $\cK(X)$ from $\cK(X')$. 

To write this more explicitly, we introduce a basis $(D_i)$ of $H^2(X)$, where $i,j,\ldots =1,\dots ,h^{1,1}(X)$, which consists of generators of the K\"ahler cone of $X$ (assumed to be simplicial). We write the K\"ahler form as $J=t^iD_i$, where $t^i>0$ corresponds to the K\"ahler cone of $X$. The relation~\eqref{eq:top_data_change} between the intersection numbers $d_{ijk}=D_i\cdot D_j\cdot D_k$ of $X$ and $d'_{ijk}$ of $X'$ can then be written as
\begin{equation}\label{isec}
 d'_{ijk}=d_{ijk}-n\,\delta_{1i}\delta_{1j} \delta_{1k}\; , 
\end{equation} 
where it is assumed that the flop arises across the boundary $t^1=0$. The integer $n$ is related to Gromov-Witten invariants and arises from the intersections with the curves $\mathcal{C}_i$ in Eq.~\eqref{eq:top_data_change}.

It has been found recently~\cite{CICYFlops, Brodie:2020fiq} that CY flop transitions are even more ubiquitous than previously thought and, in particular, that they arise for many of the simplest CY threefolds, including CICYs. In fact, of the 4874 K\"ahler-favorable CICYs, all but 6 admit a flop transition. Of these six, five have Picard number one and so could not admit a flop in principle. The remaining example is the bicubic in $\mathbb{P}^2 \times \mathbb{P}^2$.

\subsection{Symmetric flops}\noindent\label{sec:symm_flops}
Two manifolds $X$ and $X'$ which are related by a flop are (by definition) isomorphic in codimension one. But it can additionally happen that the manifolds are precisely isomorphic, that is, that the flop constitutes a birational automorphism. In particular, in our context this means that $X$ and $X'$ are isomorphic as smooth complex manifolds. We call such a flop `symmetric'.

\begin{figure}[t]
\centering
\includegraphics[height=.25\textwidth]{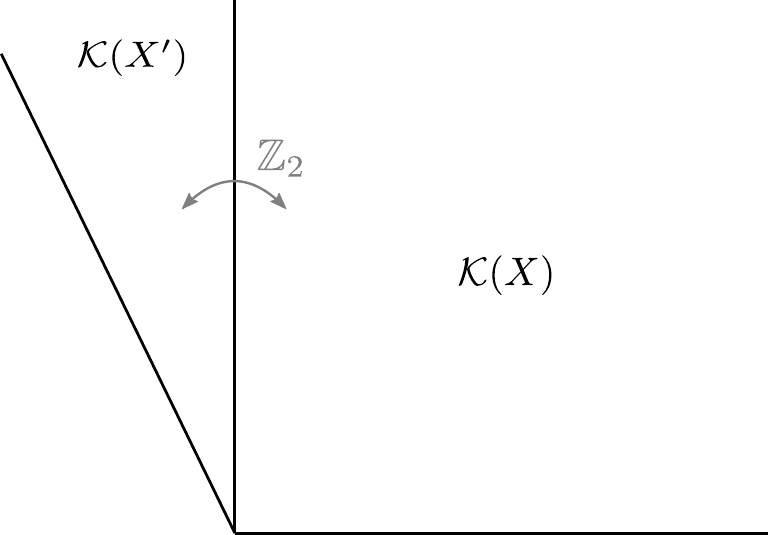}
\caption{Extended K\"ahler cone of a CY threefold admitting a symmetric flop.}
\label{fig:extendedKCsym}
\end{figure}

For a symmetric flop $X\rightarrow X'$ the moduli space has an involution $\imath$ which exchanges the K\"ahler cones $\mathcal{K}(X)$ and $\mathcal{K}(X')$, as indicated in Fig.~\ref{fig:extendedKCsym}, while leaving the boundary across which the flop arises (the vertical axis in Fig.~\ref{fig:extendedKCsym}) invariant.
Relative to the basis $(D_i)$ the involution can be represented by a matrix ${M^i}_j$ with $M^2=\mathbbm{1}$ under which the intersection numbers in Eq.~\eqref{isec} are related by a tensorial transformation, that is,
\begin{equation}\label{isecM}
 d'_{ijk}=d_{pqr}{M^p}_i{M^q}_j{M^r}_k\; .
\end{equation}
To illustrate this, it is useful to consider Picard number two manifolds. In this case, for a symmetric flop across the $t^1=0$ boundary, the involution $\imath$ is described by the matrix
\begin{equation}\label{M}
 M_1=\left(\begin{array}{cc}-1&0\\m_1&1\end{array}\right)\; .
\end{equation}
Using that $X \cong X'$ one can compute the positive integer $m_1$ in terms of the intersection numbers as $m_1=2d_{122}/d_{222}$.

Symmetric flops are quite common as well~\cite{CICYFlops, Brodie:2020fiq}. Scanning again the 4874 K\"ahler-favorable CICYs, at least 2067 admit a symmetric flop. Among the $36$ CICYs with Picard number two, $27$ display a symmetric flop and, hence, an involution $\imath$, along at least one K\"ahler cone boundary. We will discuss Picard rank 2 examples in detail in~\cite{Brodie:2021PR2}.

\subsection{Infinite flop chains}
If the manifold $X$ admits symmetric flops through multiple boundaries of its K\"ahler cone, then the extended K\"ahler cone can contain an infinite number of K\"ahler cones, connected by an infinite sequence of flops. In such cases, the corresponding involutions typically do not commute and generate a discrete symmetry of countably infinite order.

\begin{figure}[t]
\centering
\includegraphics[height=.25\textwidth]{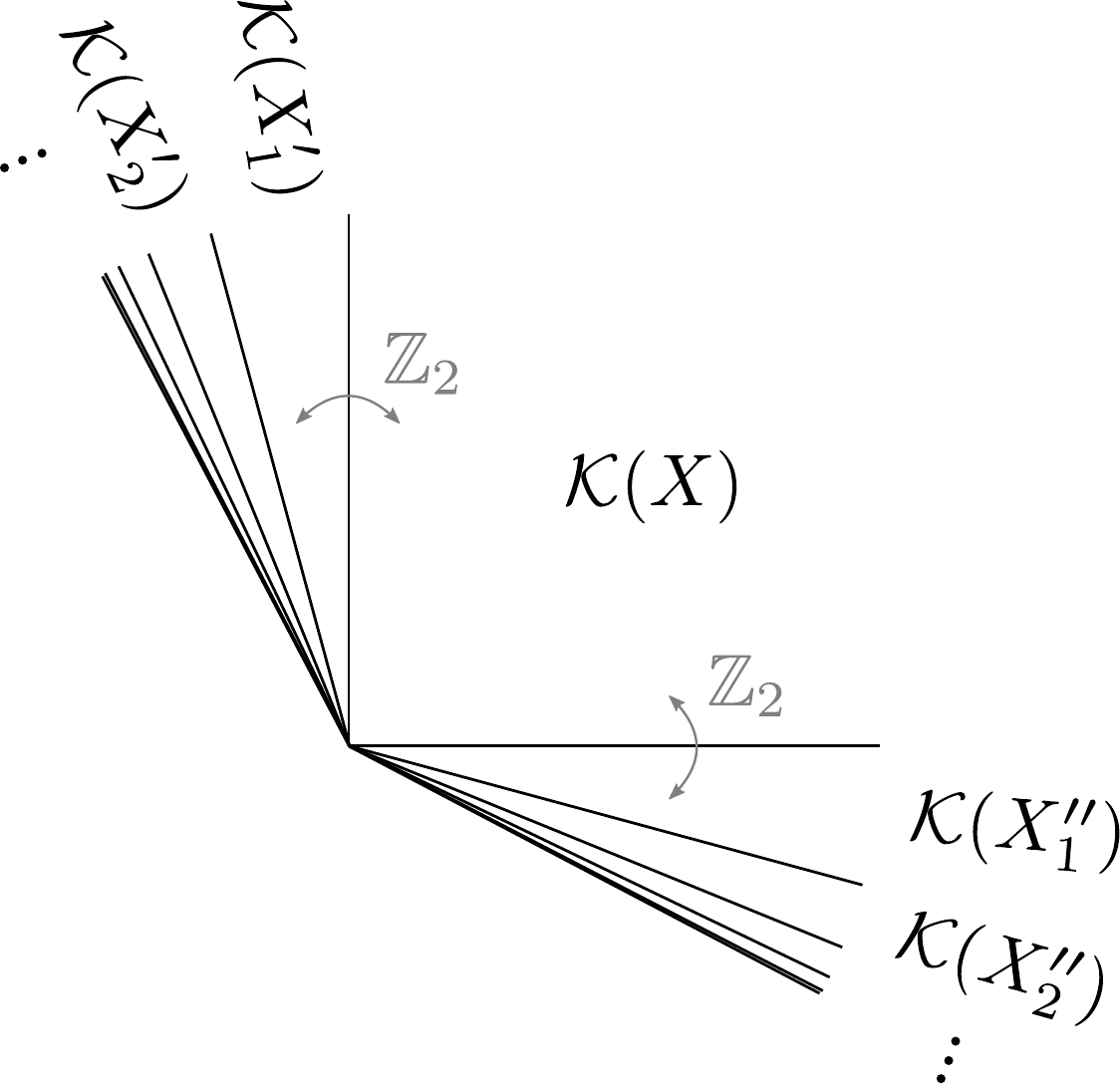}
\caption{Extended K\"ahler cone of a CY threefold admitting an infinite flop chain.}
\label{fig:extendedKCinf}
\end{figure}

The simplest situation occurs for $h^{1,1}(X)=2$, when symmetric flops arise across both boundaries of the K\"ahler cone. The resulting extended K\"ahler cone is depicted schematically in Fig.~\ref{fig:extendedKCinf}. 
In this case, $X$ admits symmetric flops to two manifolds $X'_{1}$ and $X''_{1}$. But since $X'_{1}$ is isomorphic to $X$, it must itself admit two flops, one going back to $X$, the other one going to another isomorphic manifold,~$X'_{2}$. By iteration, this process leads to an infinite number of K\"ahler cones on both sides of the original K\"ahler cone $\cK(X)$. Here the $\IZ_2$-action relating $\cK(X)$ and $\cK(X'_1)$ does not commute with the $\IZ_2$-action relating $\cK(X)$ and $\cK(X'_2)$. The two involutions can be obtained by writing down the two matrices representing $\imath_1$ and $\imath_2$ which, generalizing Eq.~\eqref{M}, are given by
\begin{equation}\label{M1}
 M_1=\left(\begin{array}{cc}-1&0\\m_1&1\end{array}\right)\; ,\quad
 M_2=\left(\begin{array}{cc}1&m_2\\0&-1\end{array}\right)\;,
\end{equation}
with $m_1=2d_{122}/d_{222}$ and $m_2=2d_{211}/d_{111}$. It is clear that $M_1^2=M_2^2=\mathbbm{1}$ but also that the two matrices do not commute. Their product
\begin{equation}
M_{12}=M_1M_2=\left(\begin{array}{cc}-1&-m_2\\m_1&-1+m_1m_2\end{array}\right)\;
\end{equation} 
is of finite order for $m_1m_2<4$ and it generates an infinite group isomorphic to $\mathbb{Z}$ for $m_1m_2\geq 4$. In fact, every element in $G$ can be uniquely written as $M_1^q M^k_{12}$, where $q\in\{0,1\}$ and $k$ is in a finite range for $m_1m_2<4$ and $k\in\mathbb{Z}$ for $m_1m_2\geq 4$. 

The three finite order cases for $m_1m_2<4$ are explicitly given in the Table below.
\begin{center}
\begin{tabular}{|c||c|c|c|}\hline
$(m_1,m_2)$&$(1,1)$&$(1,2)$&$(1,3)$\\\hline
$G\cong$&$\mathbbm{Z}_2\ltimes\mathbbm{Z}_3$&$\mathbbm{Z}_2\ltimes\mathbbm{Z}_4$&$\mathbbm{Z}_2\ltimes\mathbbm{Z}_6$\\\hline
\end{tabular}
\end{center}
For these special cases, the extended K\"ahler cone consists of a finite number of K\"ahler sub-cones, so that we have only finite sequences of flops. However, the nef cone would fill the entire plane $\mathbbm{R}^2$ (and hence not be a proper cone). Since such nef cones do not exist, CYs with two $\mathbb{Z}_2$ involutions and triple intersection numbers that would give rise to $m_1 m_2 <4$ are in the swampland.

For $m_1m_2\geq 4$, on the other hand, $G$ is isomorphic to the free product $G\cong\mathbb{Z}_2*\mathbb{Z}_2\cong\mathbb{Z}_2\ltimes\mathbb{Z}$. An infinite number of K\"ahler cones, on both sides of the original  cone $\mathcal{K}(X)$, is generated, as indicated in Fig.~\ref{fig:extendedKCinf}, and infinite sequences of flops are possible. In this case, the generators of the extended K\"ahler cone are obtained by taking the limits
\begin{equation}
v_1=\lim_{k\rightarrow\infty}\frac{M^k{\bf e}_1}{|M^ke_1|}\;,\quad
v_2=\lim_{k\rightarrow\infty}\frac{M^k{\bf e}_2}{|M^ke_2|}\; ,
\end{equation}
where $e_i$ are the standard unit vectors. Up to an irrelevant overall normalization, these vectors are given by
\begin{equation}
v_1=\left(\begin{array}{c}\tilde{m}_2\\-1\end{array}\right)\;,\quad
v_2=\left(\begin{array}{c}-1\\\tilde{m}_1\end{array}\right) \;,
\end{equation}
where $\tilde{m}_i=\frac{m_i}{2}(1+\sqrt{1-4/(m_1m_2)})$. Hence, the extended K\"ahler cone for infinite flop chains is rational in the ``critical case" $m_1m_2=4$, but it is irrational in all other cases, that is whenever $m_1m_2>4$.

Again, infinite flop chains appear to be rather common. 505 of the 4874 K\"ahler-favorable CICYs admit symmetric flops across multiple boundaries of the K\"ahler cone, and, hence, have an infinite flop chain. For the $36$ CICYs with Picard number two, $6$ have symmetric flops across both boundaries~\cite{Brodie:2021PR2}.

CICY examples of infinite flop chains appear to be such that all manifolds involved are isomorphic to one another. An obvious generalization are infinite flop chains based on a finite number, $\nu$, of non-isomorphic manifolds, for example arranged in a repeating sequence
\begin{equation}\label{seq}
 X_1\rightarrow \cdots\rightarrow X_\nu\rightarrow X_1\rightarrow\cdots\rightarrow X_\nu\rightarrow \cdots\; ,
\end{equation}
where $X_1,\ldots ,X_\nu$  are inequivalent. While such cases may exist, we are not aware of explicit examples with $\nu>1$.

A qualitatively different case is an infinite flop chain which involves an infinite number of non-isomorphic manifolds. For such cases, the Kawamata-Morrison conjecture enters our discussion.

\subsection{The Kawamata-Morrison conjecture}\noindent
In the context of the minimal model program it is important to know whether an extended K\"ahler cone can contain infinitely many non-isomorphic manifolds. For CY threefolds a negative answer has been proposed in the form of the Kawamata-Morrison conjecture \cite{1994alg.geom..7007M,Kawamata1997OnTC}.

The aim of the minimal model program is to classify irreducible complex varieties up to birational equivalence (see, for example, Chapter 12 of Ref.~\cite{Matsuki:2001}). A minimal model is a `nice' representative of a birational equivalence class, characterized by a nef canonical bundle. In particular, CYs are minimal models since their canonical bundle is trivial. Given a variety that is not a minimal model, constructing an associated minimal model involves a sequence of contractions of curves\footnote{Some of these contractions lead to varieties that are too singular, in which case flips can be used to ameliorate the singularities.} that negatively intersect the canonical divisor; eventually, the canonical divisor should become nef. Minimal models are not unique in general and any two birationally equivalent minimal models are connected by flops.

In this context a natural question is whether the number of minimal models in a birational equivalence class is always finite. For the case of complex surfaces, the answer is positive and, moreover, in a fixed birational class there is a unique minimal model. In dimension three or higher, the question is not settled. As discussed, it is not difficult to find examples of threefolds that admit an infinite sequence of flops. However, all known examples produce a finite number of minimal models up to isomorphisms. 

The Kawamata-Morrison conjecture for CY threefolds states that in a birational equivalence class there are only finitely many isomorphism classes. In other words, an infinite sequence of flops can only occur between a finite number of  isomorphism classes of CY threefolds. 

\subsection{M-theory on CY threefolds}\noindent
M-theory compactifications on CY threefolds enjoy a number of features that make the analysis of geodesic motion tractable. Notably, when we study geodesic motion in the extended K\"ahler cone in Section~\ref{sec:Geodesics}, it will be sufficient to consider only the vector multiplet moduli space, which receives no quantum corrections in the interior of the K\"ahler cone. Moreover, in contrast to four-dimensional $\mathcal{N}=1$ or $\mathcal{N}=2$ models~\cite{Palti:2020qlc}, where a rich structure of non-geometric or hybrid phases can occur~\cite{Aspinwall:2009qy,Blaszczyk:2011hs,Aspinwall:2014vea,Witten:1993yc,Aspinwall:1993xz}, the moduli space of five-dimensional $\mathcal{N}=1$ supergravity is simpler and ends at the boundary of the extended K\"ahler cone, missing all non-geometric phases~\cite{Witten:1996qb}.

The compactification of M-theory (11d supergravity) on a CY threefold $X$ was first studied in Ref.~\cite{Cadavid:1995bk}. It leads to a five-dimensional $N=1$ supergravity with a gravity multiplet, $h^{1,1}(X)-1$ vector multiplets and $h^{2,1}(X)+1$ hyper multiplets. The overall volume modulus $V={\rm vol}(X)$ is part of the hyper multiplet sector while the relative K\"ahler moduli $b^i=t^i/V$ form the scalars within the vector multiplets $(b^i,A^i,\lambda^i)$, where $A^i$ are the Abelian gauge fields and $\lambda^i$ the gauginos. Since the $b^i$ are subject to the constant volume constraint
\begin{equation}\label{kappa}
 \kappa:=d_{ijk}b^ib^jb^k\stackrel{!}{=}6
\end{equation}
we are indeed left with $h^{1,1}(X)-1$ independent scalar fields, as required\footnote{Of the $h^{1,1}(X)$ vector fields and fermions one combination each becomes part of the gravity multiplet.}.

The hyper multiplet scalars form a quaternionic geometry which receives loop corrections, such as the one computed in Ref.~\cite{Strominger:1997eb}. The vector multiplet scalars, on the other hand, parametrize a manifold with very special geometry, governed by the tri-linear pre-potential~\eqref{kappa}. Since the volume modulus $V$ is part of the hyper multiplet sector, the vector multiplet geometry can be computed at large volume and does not receive corrections. In our context, we will be primarily interested in the vector multiplet moduli $b^i$ and their associated geodesics.

Besides the Hodge numbers, the five-dimensional supergravity is determined by the triple intersection numbers $d_{ijk}$ of the underlying CY threefold.  In particular, the moduli space metric for fields $b^i$ is given in terms of the prepotential~\eqref{kappa} as
\begin{equation}
 G_{ij}=-\frac13\partial_i\partial_j\ln\kappa=-2\left(\frac{\kappa_{ij}}{\kappa}-\frac32\frac{\kappa_i\kappa_j}{\kappa^2}\right)\; ,
\end{equation}
where $\kappa_i=d_{ijk}t^jt^k$ and $\kappa_{ij}= d_{ijk} t^k$. The resulting geodesic equation (decoupling five-dimensional gravity) is
\begin{equation}\label{geodesic}
  \ddot{b}^i+\Gamma^i_{jk}\dot{b}^j\dot{b}^k=0\,,
\end{equation}
where the dot denotes the derivative, $d/ds$, with respect to an affine parameter $s$ parameterizing the geodesic curve and $\Gamma^i_{jk}=\frac{1}{2}G^{il}\partial_l G_{jk}$  is the connection. This equation should be solved subject to the constant volume constraint~\eqref{kappa}.

Using standard relations of very special geometry it is straightforward to show that the geodesic equation~\eqref{geodesic} has the first integral
\begin{equation}
 \frac{1}{2}G_{ij}\dot{b}^i\dot{b}^j=E\; ,
\end{equation}
where $E$ is a non-negative constant. This means the geodesic distance $\Delta\tau$ of a path $b^i(s)$ with $s\in[s_1,s_2]$ and $\Delta s=s_2-s_1$ can be computed as
\begin{equation}
\label{eq:GeodesicDistanceKahler}
\Delta\tau=\int_{s_1}^{s_2}ds\,\sqrt{\frac{1}{2}G_{ij}(b(s))\dot{b}^i\dot{b}^j}=\sqrt{E} \, \Delta s\; .
\end{equation}
This provides us with the relevant ingredients of the five-dimensional theory as long as the moduli $b^i$ remain in the interior of the K\"ahler cone.

What happens when a flop boundary, say $b^1=0$, is approached? First, such a flop boundary can be reached in a finite geodesic distance. Further, at the boundary we have to consider a number of additional hyper multiplets which originate from membranes wrapping the shrinking curves $\mathcal{C}_i$. Their mass is proportional to $b^1$, so they become massless at the transition and will be referred to as transition states. A five-dimensional effective theory including the transition states has been developed in Refs.~\cite{Brandle:2002fa,Jarv:2003qx,Jarv:2003qy}. Ref.~\cite{Witten:1993yc} shows that the transition states lead to one-loop corrections which change the intersection numbers $d_{ijk}$ of $X$ to the intersection numbers $d'_{ijk}$ of $X'$, as given in Eq.~\eqref{isec}, when the flop boundary $b^1=0$ is crossed. This means that we can think of the five-dimensional theory as a theory on the extended K\"ahler cone, as long as we adapt the intersection numbers to the K\"ahler cone under consideration.

More concretely, suppose we would like to consider a geodesic $b^i(s)$ which is contained in the K\"ahler cone $\mathcal{K}(X)$ for $s<0$, crosses the flop boundary at $s=0$ and extends into the K\"ahler cone $\mathcal{K}(X')$ for $s>0$. From the above discussion, this geodesic should satisfy Eq.~\eqref{geodesic} for $s<0$ and for $s>0$ it should satisfy the same equation but with the intersection numbers $d_{ijk}$ in Eq.~\eqref{kappa} (as well as in $G_{ij}$ and $\Gamma^i_{jk})$ replaced by $d'_{ijk}$. At the flop transition we should require continuity of $b^i(s)$ and $\dot{b}^i(s)$.

We can say a bit more if the flop $X\rightarrow X'$ under consideration is symmetric, that is, if $X$ and $X'$ are isomorphic CY threefolds and are related by an involution described by a matrix $M$. In this case, the intersection numbers are related as in Eq.~\eqref{isecM} and the involution becomes a symmetry of the five-dimensional supergravity (understood as a theory on the extended K\"ahler cone, as discussed above) which acts as
\begin{equation}\label{invb}
 b^i\rightarrow {M^i}_jb^j\; ,
\end{equation}
and similarly on the gauge fields $A^i$ and the gauginos $\lambda^i$, while leaving all other fields invariant. 

\begin{figure}[t]
\centering
\includegraphics[width=.24\textwidth]{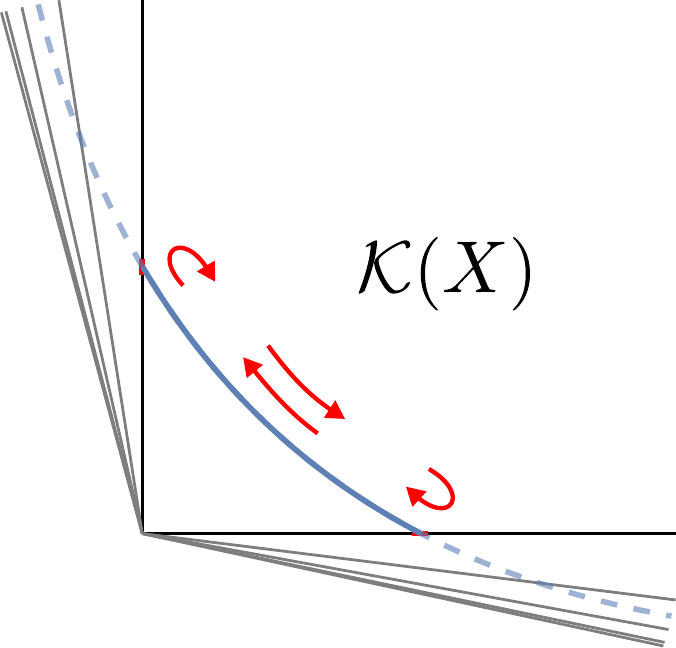}
\caption{The mapping (in blue) into a single K\"ahler cone $\cK(X)$ of a constant volume geodesic (dashed) which traverses an infinite number of K\"ahler cones connected by symmetric flops.}
\label{fig:oscil}
\end{figure}

We can apply this symmetry to the $s>0$ part of the geodesic (which is within $\mathcal{K}(X')$) to map it back into the K\"ahler cone $\mathcal{K}(X)$, i.e.\ ${M^i}_jb^j(s)$ with $s>0$ solves the original geodesic equation~\eqref{geodesic} with intersection numbers $d_{ijk}$. Moreover, since we have imposed continuity of $b^i(s)$ and $\dot{b}^i(s)$ across the flop and the flop boundary is invariant under $M$, the geodesics $b^i(-s)$ and ${M^i}_jb^j(s)$ for $s>0$ have the same initial conditions. Hence, they must be identical, that is,
\begin{equation}\label{bounce}
 {M^i}_jb^j(s)=b^i(-s)\quad\mbox{for}\quad s\geq 0\; .
 \end{equation}
This observation allows us to describe the geodesic motion across a flop within a single K\"ahler cone $\mathcal{K}(X)$. Instead of crossing the boundary $b^1=0$ into the K\"ahler cone $\mathcal{K}(X')$ we can instead think about the motion ``bouncing back" from the flop boundary and, in accordance with Eq.~\eqref{bounce}, retrace its original path, as indicated in Fig.~\ref{fig:oscil}. We will discuss this in detail for the Picard rank~2 cases in Ref.~\cite{Brodie:2021PR2}. 

If a symmetric flop happens at two K\"ahler cone boundaries so that we have an infinite flop chain, we can have geodesics traversing an arbitrary number of K\"ahler cones. However, now the five-dimensional supergravity has an infinite discrete symmetry group $G$ whose elements $M$ act as in Eq.~\eqref{invb}. This symmetry can be used to map the geodesic in each K\"ahler cone in Fig.~\ref{fig:extendedKCinf} back into $\mathcal{K}(X)$, so that the entire geodesic can be thought of as an oscillatory motion between the two flop boundaries of $\mathcal{K}(X)$. For Picard number two this is indicated in Fig.~\ref{fig:oscil} (see also Ref.~\cite{Brodie:2021PR2}).

As an aside, we remark that the symmetry $G$ of the five-dimensional theory might well be broken when other elements are added to the compactification. For example, a flux compactification with four-form flux on $X$ adds gauging and a potential to the five-dimensional theory. In compactifications of heterotic M-theory to five dimensions~\cite{Lukas:1998tt} such a flux is automatic and related to the second Chern classes of the CY manifold and the heterotic vector bundles. It is at present unclear how four-form flux or, in the heterotic case, vector bundles map across a flop transition and whether this process does or does not preserve the symmetry $G$. Also, in the presence of a potential on the moduli space, a flop can only be achieved if the flop boundary can be reached along flat directions of the potential. Whether or not this is possible depends on the example under consideration.

\section{Flopping geodesics and the swampland distance conjecture}\label{sec:Geodesics}
\noindent
Now that we have described the main characteristics of flopping geodesics, we are ready to discuss the interplay with the swampland distance conjecture. First, we recall that the five-dimensional massless spectrum is determined by the CY Hodge numbers. Since these remain unchanged under any kind of flop (symmetric or not) a flopping geodesic does not affect the five-dimensional massless spectrum. (The transition states only become massless at the flop locus and are massive in the interior of any K\"ahler cone.) We begin by discussing the case of a single non-symmetric flop, followed by a discussion of a single symmetric flop, a discussion of infinitely many symmetric flops, and finally infinitely many non-symmetric flops.

\begin{figure*}[t]
\centering
\includegraphics[height=.2\textwidth]{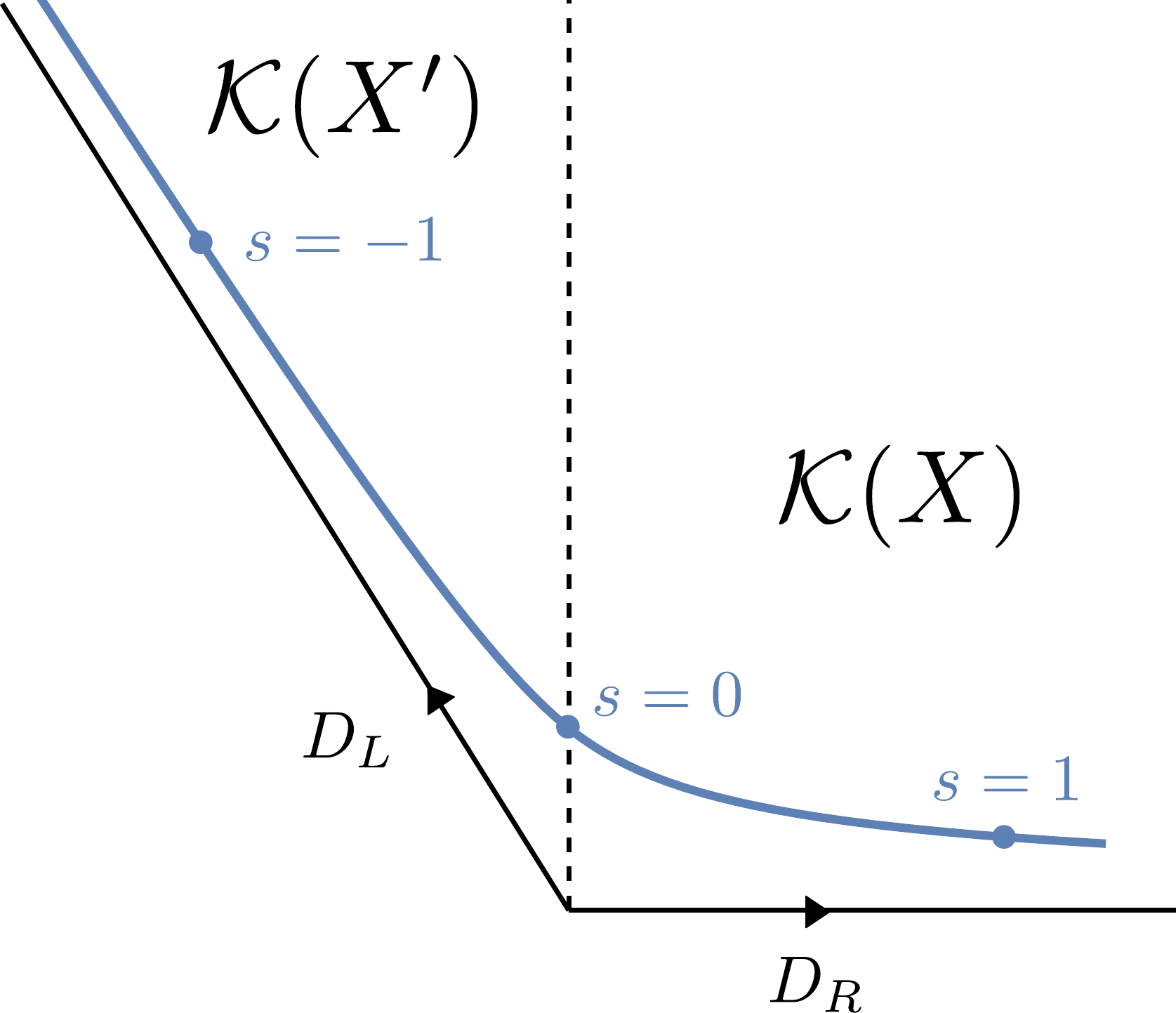}\hspace{4cm}
\includegraphics[height=.2\textwidth]{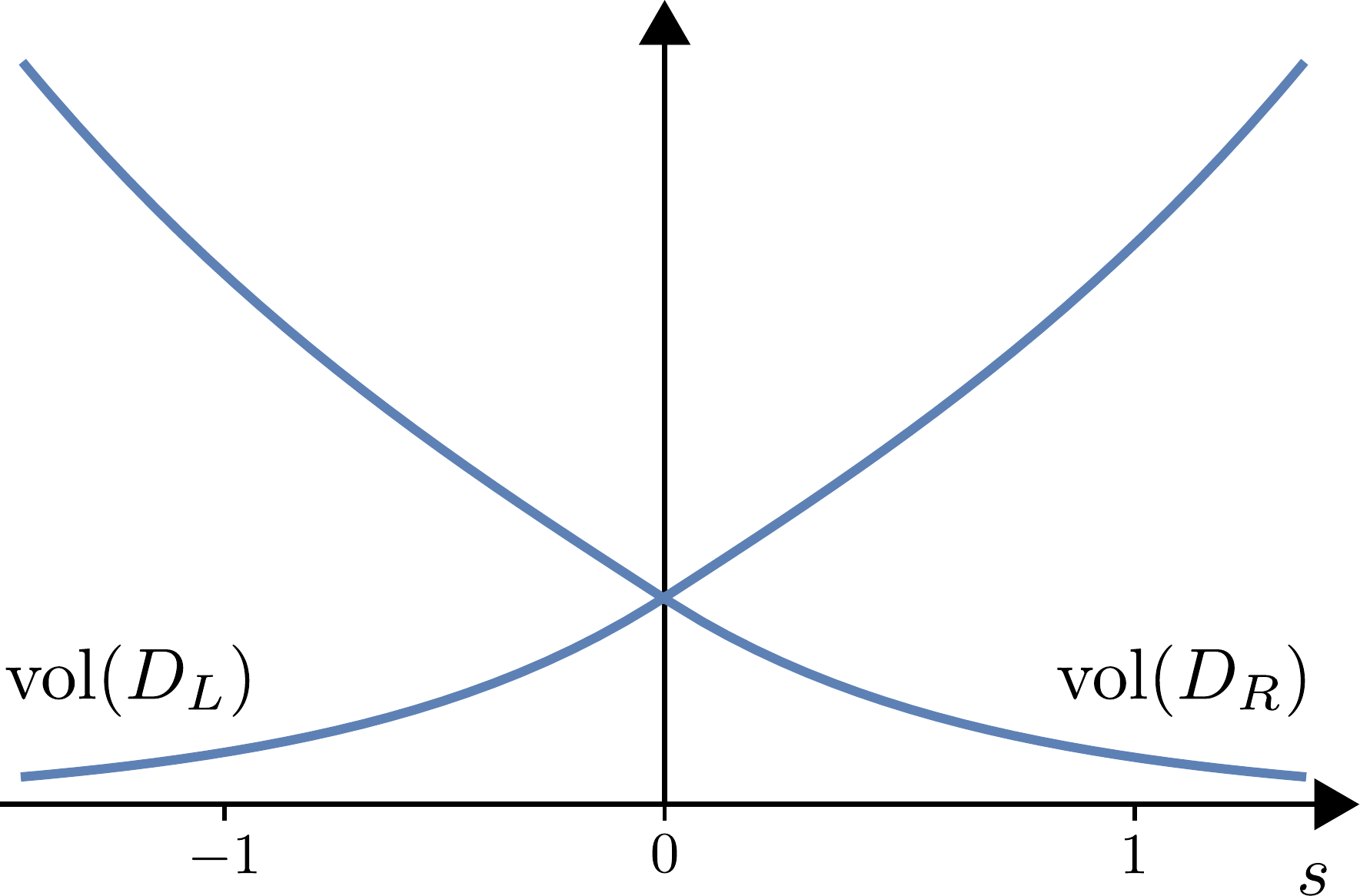}
\caption{(Left): Constant-volume geodesic across a flop boundary. (Right): Behavior of the volumes of the boundary divisors $D_L$ and $D_R$ during the geodesic motion.}
\label{fig:volumes}
\end{figure*}

\subsection{Geodesics across non-symmetric flops}\label{sec:nonsymm}\noindent
Suppose a geodesic $b^i(s)$ evolves across a non-symmetric flop $X\rightarrow X'$ from a point $p_0$ in the interior of $\mathcal{K}(X)$ to a point $p_1$ in the interior of $\mathcal{K}(X')$. The geodesic length $\Delta\tau$ is non-zero but finite and the massless spectrum at the start and end point is identical. At $p_1$ we expect a tower of light states starting at $e^{-\rho\Delta\tau}$, compared to $p_0$ (and vice versa). Checking this explicitly would typically require calculating the massive spectrum for CY compactifications as a function of the moduli $b^i$ as, for example, was done in Ref.~\cite{Ashmore:2021qdf}. Such calculations are beyond the scope of this paper.

The existence of a candidate tower of states is clear in the situation when the effective cone contains only two K\"ahler cones, $\mathcal{K}(X)$ and $\mathcal{K}(X')$, as shown in Fig.~\ref{fig:volumes}. Approaching the boundaries of the effective cone corresponds in this situation to a divisor shrinking down to a point, and a wrapped M5-brane gives rise to an infinite tower of states. The corresponding geodesic is shown in Fig.~\ref{fig:volumes} (left), and the volumes of the two boundary divisors are plotted in Fig.~\ref{fig:volumes} (right), making clear the exchange of two towers of massive states in moving between $p_0$ to $p_1$.

\subsection{Geodesics across symmetric flops}\noindent
Next we consider a symmetric flop $X\rightarrow X'$ with associated involution generated by $M$. Suppose a geodesic $b^i(s)$ evolves from a point $p_0\in\mathcal{K}(X)$ across the flop to the \textit{equivalent} point $p_1=Mp_0\in\mathcal{K}(X')$. Necessarily, the (massless and massive) spectra at $p_0$ and $p_1$ are precisely identical. Naively this appears to be in tension with the distance conjecture.

While a picture analogous to Fig.~\ref{fig:volumes} emerges, which would also save the conjecture for finitely many symmetric flops, the resolution in this case is a different one (and it will also apply to the infinite symmetric flop chain). Note that the $\IZ_2$-symmetry relating $\cK(X)$ and $\cK(X')$ translates into a global $\IZ_2$-symmetry of the 5d supergravity theory. Since in quantum gravity all symmetries are conjectured to be either gauged or broken~\cite{Misner:1957mt, Banks:2010zn}, this $\IZ_2$-symmetry must be gauged. Hence the moduli space is actually not the extended K\"ahler cone $\mathcal{K}_{\rm ext}(X)=\mathcal{K}(X)\cup\mathcal{K}(X')$ but its quotient $\mathcal{K}_{\rm ext}(X)/\mathbb{Z}_2$ whose interior is isomorphic to $\mathcal{K}(X)$. Hence, the points $p_0$ and $p_1$ must strictly be identified, and the shortest geodesic between $p_0$ and $p_1$ is the constant geodesic $b(s)=p_0$, rather than the one we have originally considered. Hence, there is no tension with the swampland distance conjecture . Note also that consequently the bounce description of geodesic flops discussed in the previous section is the physically correct picture.

What is the origin of the gauging of the $\IZ_2$-symmetry? The answer is clear when the problem is phrased in the right language. As discussed in Section~\ref{sec:symm_flops}, the threefolds $X$ and $X'$ are isomorphic as smooth complex manifolds. In particular, this means that $X$ and $X'$ are related by a diffeomorphism, which moreover maps a K\"ahler form $J\in\cK(X)$ to the corresponding K\"ahler form $MJ\in\cK(X')$. As such, the discrete global symmetries arising in compactifications on manifolds admitting symmetric flops are part of the diffeomorphism group of the compactification space, and are hence remnants of the gauged diffeomorphism symmetry in 10d or 11d.

\subsection{Geodesics across infinite flop chains}\noindent 
Generalizing the discussion from the previous subsection, we can consider an infinite flop chain $X=X_0\rightarrow X_1\rightarrow X_2\rightarrow \cdots$ of isomorphic CYs $X_a$, and a geodesic $b^i(s)$ which moves across an arbitrary number of these cones and connects equivalent points $p_a\in\mathcal{K}(X_a)$.

In this case, the total geodesic length $\Delta\tau$ is unbounded while the massless and massive spectrum is the same at each point $p_a$. Further, unlike the situation for a finite number of symmetric flops, depicted in Fig.~\ref{fig:volumes}, there are no boundary divisors at the end of the extended K\"ahler cone, and hence there is no exchange of associated towers of massive states along the geodesic.

The five-dimensional theory is invariant under the infinite order discrete symmetry $G$ and, from the above discussion, we should think of $G$ as a gauge symmetry. Hence, the proper moduli space is $\mathcal{K}_{\rm ext}(X)/G$ and the interior of this space is isomorphic to $\mathcal{K}(X)$. All the points $p_a$ are identified under $G$ and the geodesic under discussion is, hence, no longer the shortest available. As before, a conflict with the distance conjecture is avoided. It is also worth noting that the original geodesic $b^i(s)$, as seen in the quotient moduli space $\mathcal{K}_{\rm ext}(X)/G$, is described by the oscillatory motion indicated in Fig.~\ref{fig:oscil}.

We can also consider an infinite flop chain which contains a finite number of non-isomorphic CYs $X_1,\ldots  X_\nu$, for example forming a repeating sequence~\eqref{seq}. In this case, taking the quotient $\mathcal{K}_{\rm ext}(X)/G$ will reduce the extended K\"ahler cone to a finite number of K\"ahler sub-cones, $\mathcal{K}(X_1),\ldots ,\mathcal{K}(X_\nu)$. 
As in the above case of $\nu = 1$, after the identification there is a bound on the shortest geodesic between any two points at constant volume. However, checking consistency of the finite length geodesics with the distance conjecture would require calculating the massive spectrum for this compactification.

\subsection{Infinitely many non-isomorphic flops}\noindent 
Finally, consider an infinite sequence of flops $X_1\rightarrow X_2\rightarrow \cdots$ with all $X_a$ (or an infinite number of them) not isomorphic. Consider a geodesic $b^i(s)$ which traverses an arbitrary number of K\"ahler cones $\mathcal{K}(X_a)$, with a geodesic distance $\Delta \tau_a$ in each. It might well be the case that the total geodesic distance, obtained by summing the $\Delta\tau_a$, is unbounded. However, unlike in the case of isomorphic CYs, there is no recourse to gauge symmetries to bound the length of the geodesics. Then, the distance conjecture implies the eventual appearance of (effectively) massless states, while we know that the massless spectrum is identical for each~$X_a$.

This potential conflict with the distance conjecture can be avoided if we assume that the Kawamata-Morrison conjecture is true. In this case, flop chains containing an infinite number of non-isomorphic CYs are expressly excluded. Conversely, the Kawamata-Morrison conjecture for CY threefolds follows from applying the swampland distance conjecture to infinite flop chains, provided that the geodesics traversing infinitely many K\"ahler cones indeed have infinite length.

\section{Conclusion}\label{sec:conclusion}
\noindent 
In this letter, we have discussed the relation between topological transitions, specifically flops, in string theory and the swampland distance conjecture. 

We were motivated by recent results~\cite{CICYFlops, Brodie:2020fiq} which show that infinite chains of CY flops are not only possible but are, in fact, a common feature of relatively simple constructions of CY manifolds. This suggests the existence of geodesics across an arbitrary number of K\"ahler cones with arbitrary lengths but an essentially unchanged spectrum, in tension with the swampland distance conjecture. 

To investigate this problem we have studied geodesics in the vector moduli space of five-dimensional $N=1$ supergravity theories obtained from M-theory compactifications on CY threefolds. We have found that the potentially problematic geodesics across several --- or indeed an arbitrary number of --- K\"ahler cones, do exist. However, we argue that in each case there is a way out of a conflict with the distance conjecture, although the nature of the resolution depends on the structure of the extended K\"ahler cone.

For a flop $X\rightarrow X'$ between two non-isomorphic CYs, finite but non-zero length geodesics connecting a point $p_0\in\mathcal{K}(X)$ with a point $p_1\in\mathcal{K}(X')$ exist. However, while the initial and final spectra may be similar, the flop transition exchanges distinct towers of massive states, leading to consistency with the distance conjecture.

For flops $X\rightarrow X'$ between isomorphic CYs, there exist geodesics which connect a point $p_0\in\mathcal{K}(X)$ with the equivalent point $p_1\in\mathcal{K}(X')$ having identical massless and massive spectrum. For such cases we find a gauged $\mathbb{Z}_2$-symmetry which identifies the cones $\mathcal{K}(X)$ and $\mathcal{K}(X')$ and the points $p_0$ and $p_1$ in particular. Hence, the shortest path between $p_0$ and $p_0=p_1$ corresponds to not moving at all, trivially satisfying the distance conjecture.

For infinite flop chains $X\rightarrow X_1\rightarrow X_2\rightarrow \cdots$ of isomorphic CYs $X_a$, the moduli space is obtained by taking the quotient with an infinite discrete gauge symmetry $G$ reducing it effectively to a single  K\"ahler cone $\mathcal{K}(X)$. Thus, geodesics traversing arbitrarily many K\"ahler cones, which can reach infinite length without the appearance of a tower of states, are no longer the shortest ones and are, hence, unproblematic. As a by-product we arrive at a surprising picture for geodesic motion across symmetric flops. We can think of such geodesics as ``bouncing back" from the flop boundary or, in the case of two flop boundaries, as oscillating between them (see Fig.~\ref{fig:oscil}).

Infinite flop sequences $X_1\rightarrow X_2\rightarrow\cdots$ of non-isomorphic CYs might well constitute a problem for the distance conjecture. However, such sequences are excluded, provided the Kawamata-Morrison conjecture holds. Turning the argument around, the Kawamata-Morrison conjecture is implied by the swampland distance conjecture, provided that constant-volume geodesics of arbitrary lengths can be constructed.

\begin{acknowledgments}
We thank Ben Heidenreich and Tom Rudelius for useful discussions. The work of CRB is supported by the John Templeton Foundation grant 61149. AC would like to thank EPSRC for grant EP/T016280/1.
\end{acknowledgments}

\bibliography{refs}

\end{document}